\documentstyle[prd,aps,epsfig,preprint]{revtex}
\tighten
\begin{document}
\thispagestyle{empty}
\baselineskip24pt
\draft
\begin{center}
{\large\bf The gauge freedoms of enlarged Helmholtz theorem and the Neumann --- Debye
potentials; their manifestation in the multipole expansion of conserved 
current}
\end{center}
 \begin{center}
{Elena~N.~Bukina and Vladimir~M.~Dubovik}\\
{\it Joint Institute for Nuclear Research,
          141980 Dubna, Moscow region, Russia} \\
{\it e-mail: bukina@thsun1.jinr.ru, dubovik@thsun1.jinr.ru} 
\end{center}
\date{\today}
\begin{abstract}
We discuss gauge freedom within the scope of the enlarged Helmholtz 
theorem and Neumann-Debye decomposition and then demonstrate its realization
for the multipole expansion of a electromagnetic current with distinguished 
toroid moment family.
The exact solution to the latter problem was obtained in 1974,
but answers to some purely mathematical questions
raised by I.B.~French and Y.~Shimamoto~\cite{French} about 40 years ago 
are given in this paper for the first time. 
\end{abstract}

\section{Introduction}

$\quad$In considering problems of mathematical physics with definite spatial
and/or dynamical symmetries, one commonly uses various decompositions
of vector fields over scalar potentials. These decompositions supplement
the famous Helmholtz theorem and reduce its ``gauge freedoms''.

Let $\exists \ \mbox{\bf V}: \ \mbox{\bf r} \in R^3  \rightarrow \mbox{\bf V}(\mbox{\bf r})
\in  R^3 $ with all ``good'' properties. We may represent the given field
$\mbox{\bf V}$ in different ways

$$\mbox{\bf V} \rightarrow \{ V_x, V_y, V_z\} \rightarrow \{ \varphi ,
\mbox{\bf A} \}_{\mbox{div} \mbox{\bf A}=0} \rightarrow \{\varphi , \psi , \chi\}.$$
In any case, it is necessary to constrain superfluous components,  if they
take place, by introducing conditions similar to $\mbox{div} \mbox{\bf A}=0.$
                             
The last variant, diffeomorphic scalarization of a vector field, is the most economical and 
convenient approach to vector boundary-value problems of mathematical physics.
But to use this approach, we have to be able to invert  
decomposition formulas, i.e. to deduce the integral representations of scalar 
potentials through the original vector field.

{\large Approach.}
To obtain a vector function from a (pseudo)scalar function set
$\varphi(\mbox{\bf r}), \psi(\mbox{\bf r})$ and $\chi(\mbox{\bf r})$, one must 
act on them by some vector
operator $\hat F(\mbox{\bf r}, \mbox{\boldmath $\nabla$} ).$  Besides $\mbox{\bf r} \ \mbox {and} \
\mbox{\boldmath $\nabla$} $ themselves, we may from them construct three simplest operators:
$\ \mbox{\bf L}:=-\mbox{\bf r} \times \mbox{\boldmath $\nabla$}, \ 
\mbox{\bf N}:= \mbox{\boldmath $\nabla$} \times
\mbox{\bf L} \ \mbox {and} \ \mbox{\bf M}:=-\mbox{\bf r} \times \mbox{\bf L}.$ One can see their
symplectic nature because of their correspondence to the frames of reference
in a phase space: $\mbox{\bf r},\ \mbox{\bf k},\  \mbox{\bf r} \times \mbox{\bf k}, $
$\mbox{\bf r}\times (\mbox{\bf r}\times \mbox{\bf k})$
and $\mbox{\bf k} \times (\mbox{\bf k} \times \mbox{\bf r}).$ 
The {\it trio} of vectors $\mbox{\bf k},\ \mbox{\bf L},\ \mbox{\bf N}$ and $\mbox{\bf r},\ \mbox{\bf L},\ \mbox{\bf N}$, being
immersed into the spaces $R_{\mbox{\bf r}}^{3}$ and $R^{3}_{\mbox{\bf p}}$, respectively, form orthogonal
bases in them, which are important for different applications.

One may verify the 
following projection and commutation properties of $\hat F$:\\

$[\mbox{\bf L}, r^2] = [\mbox{\bf L}, p^2] = 0$

$[L^2, r] = i [\mbox{\bf r} \times \mbox{\bf L}] - i[\mbox{\bf L} \times \mbox{\bf r}]$

$[L^2, k] = i [\mbox{\bf k} \times \mbox{\bf L}] - i[\mbox{\bf L} \times \mbox{\bf k}]$

$\mbox{\bf r} \cdot \mbox{\bf L} = \mbox{\boldmath $\nabla$} 
\mbox{\bf L} = \mbox{\bf r} \cdot \mbox{\bf N} =
\mbox{\bf L} \cdot \mbox{\bf N} = \mbox{\bf L} \cdot \mbox{\bf M} = \mbox{\bf M} \cdot \mbox{\bf L} = 0,
\mbox{curl} \mbox{\bf N} = -\mbox{\bf L} \triangle; $

$[\mbox{\bf L}, \triangle ]=0,\  [\mbox{\bf N}, \triangle ]=0,\
[\mbox{\bf M}, \triangle ]=-6 \mbox{\boldmath $\nabla$},$ etc.

$[r_i, \nabla_k]=-\delta_{ik},\ \ [r_i,L_k]=-\epsilon_{ikj}r_j, \ \
[\nabla_i,L_k]=-\epsilon_{ikj} \nabla_j,$

$[\nabla_i,N_k]=\nabla_i\nabla_k-\triangle\delta_{ik}, \ \
[r_i,M_k]=r_ir_k-r^2\delta_{ik}$

$[L_i,M_k]=\epsilon_{ikj}r_j - r^2\epsilon_{ikj}\nabla_j.$

$..................................................................$

It is taken into account that
\begin{eqnarray}
\mbox{\bf N}=-\mbox{\bf r}\triangle + \mbox{\boldmath $\nabla$}
(\mbox{\bf r}\mbox{\boldmath $\nabla$}) + \mbox{\boldmath $\nabla$}, 
\nonumber \\ \mbox{\bf r}\times \mbox{\bf N} = -(1+\mbox{\bf r}
\mbox{\boldmath $\nabla$})\mbox{\bf L}=-\mbox{\bf L}(1+\mbox{\bf r}
\mbox{\boldmath $\nabla$}), \nonumber \\ \mbox{curl} \mbox{\bf N}=-\mbox{\bf L}
\triangle, \ \mbox{etc.} \nonumber
\end{eqnarray}

Note that in the space $R_{\mbox{\bf r}}^{3}$, $\mbox{\bf L}$ and $\mbox{\bf N}$ form the following 
algebra of differential operators
$$[L_i, L_j]=\epsilon _{ijk}L_k,\ [L_i,N_j]=\epsilon _{ijk}N_k,\ [N_i,N_j]=
-\epsilon _{ijk}L_k \triangle .$$
After the rescaling of $\mbox{\bf N} \rightarrow \mbox{\bf N}/ \sqrt{\triangle}$ over a
supporting function space, we acquire the $O(3,1)$ Lie algebra representation
in terms of the $\hat F(\mbox{\bf r}, \mbox{\boldmath $\nabla$} )$ operators in $R_{\mbox{\bf r}}^{3}$. They act in the
$O(3)\times O(3)$ function space suitable for arranging  the multipole
phenomenology in electromagnetic theory (see e.g. \cite{Bl}, \cite{DbCh}, 
\cite{DT}). 
The operators $\mbox{\bf L}$ and $\mbox{\bf M}$ obey the same algebra in the space of wave 
vectors $R_{\mbox{\bf k}}^{3}$ that is Fourier-conjugate to $R_{\mbox{\bf r}}^{3}$.
In the preceding paper \cite{DM} the emphasis has been made on the inversion of
different formulas for decomposition of vector fields in the mathematical
aspect. 

In the physical aspect, two vector-potential formulations of the theory 
of continuous media with taking 
into account both magnetic and electric toroid polarizations \cite{DTT},
(see also \cite{DubTug}, part II) are published for the first time.
Here we consider the mathematical underlying reason of uniqueness of 
division of the transverse electric distribution density
$E_{lm}(\mbox{\bf k}^{2}, \, t)$ into two independent multipole specimen~\cite{DbCh}:
\begin{equation}
E_{lm}(\mbox{\bf k}^{2}, \, t) = \dot{Q}_{lm}(0, \, t) + \mbox{\bf k}^2 T_{lm} (\mbox{\bf k}^{2}, \, t),
\label{N1}
\end{equation}
where $\dot{Q}_{lm}(0, \, t)$ are the time-derivative of the 
Coulomb (charge) multipole $l-$moment and $T_{lm} (\mbox{\bf k}^{2}, \, t)$ are
the toroid multipole form factor of $l-$order. Note that a secularized 
relation when neglecting toroid contributions and known as the Siegert theorem
may be correct for low-symmetric electromagnetic systems only.
The mathematical question of condition responsible for the possibility of
identification of multipole moments (the leading ones for each given $l$) 
in the transverse and longitudinal
parts of current raised in \cite{French} will be tried to clear in this paper.
From our representation theorem (see also remark
II) it follows that definitions of multipole moments are unique.

Recall that the procedure of multipole
decomposition of the current and field densities in electrodynamics in fact
corresponds to the description of properties of a system by a set of numerical
characteristics which are assigned to a point, ``center'' of the density
distribution of the system considered. 
In this case, for the poloidal and potential parts of 
the current an additional connection  arises between its longitudinal and
transverse components \cite{DbCh} due to degeneracy of boundary conditions
of longitudinality and transversality in $\mbox{\bf r}$-space at the self-similar
shrinkage of the definition domain of current to the chosen center. 
To prove this statement, we could probably use the transfer technique of 
boundary conditions \cite{Vlad}, \cite{Fedor}.
It might have lead to the separation of multipole moments 
called in \cite{DbCh} the toroid ones~\footnote{In fact, their exact 
title should be the poloidal moments, see~\cite{DT}, \cite{DubTug}.}. 
But ways of this kind is very difficult in the general framework of 
the distribution theory. Here we use the simple concrete approach.

The main feature of the multipole expansion procedure is a
special choice of basis functions, which in actual practice ensues the 
rapid convergence of multipole series.
This circumstance forces us to weaken requirements of the usual 
Helmholtz theorem and, respectively, to take account of a
gauge freedom extension.
In sections 1 and 2, we discuss gauge freedoms in the Helmholtz
and Neumann-Debye decomposition. In section 3, we turn to their realizations
within the multipole expansion of the electromagnetic current.


\section{The enlarged Helmholtz theorem}

$\quad$We begin our consideration with the Helmholtz Decomposition:\\
$\triangleleft \qquad \forall \mbox{\bf V}$ with properties of single-valuedness,
continuity, boundedness or convergence
$$|\mbox{\bf V}|<\frac{k}{r^{2+\epsilon}},\ \epsilon>0 \ \mbox{at}\ r\rightarrow
{\infty}$$
in the space may be represented in the form [see e.g. \cite{Lee}]:

$$\mbox{\bf V}=\mbox{\boldmath $\nabla$} \varphi +
 \mbox{curl}\mbox{\bf A} \qquad
\mbox{with} \qquad \mbox{div}\mbox{\bf A}=0.$$
Indeed, the theorem requirements are proved to be sufficient in order to
reexpress tautologically the given vector field $\mbox{\bf V}$ through its 
divergence and vorticity. The explicit realization of the theorem could be 
attained due to the following operations
\begin{eqnarray}
\mbox{div} \mbox{\bf V}^{\|}&=&\triangle \varphi, 
\varphi=\triangle ^{-1}\mbox{div}\mbox{\bf V},  \nonumber \\
\mbox{where} \quad
\triangle^{-1}&:=&-\int_{\Omega \subseteq R^3/\{0\}}{\frac{d^3r'}{4\pi |\mbox{\bf r}-
\mbox{\bf r}'|}}, \nonumber\\
\mbox{curl} \mbox{\bf V}^{\bot}&=&\mbox{curl}\ \mbox{curl}\mbox{\bf A}=-\triangle \mbox{\bf A}, \quad
\mbox{\bf A}=-\triangle^{-1}\mbox{curl}\mbox{\bf V}.\nonumber
\end{eqnarray}
So we have
\begin{equation}
\mbox{\bf V}\equiv \mbox{\bf V}^{\|}+\mbox{\bf V}^{\bot}=\mbox{\boldmath $\nabla$}
 \triangle^{-1}\mbox{div} \mbox{\bf V}
-\mbox{curl}\triangle^{-1}\mbox{curl}\mbox{\bf V}. \qquad \triangleright
\label{1}
\end{equation}
Form (\ref{N1}) often produces misunderstanding
(e.g. \cite{Bosko}) that the representation of $\mbox{div} \mbox{\bf V}$ and
$\mbox{curl} \mbox{\bf V}$ is equivalent to the representation of $\mbox{\bf V}$ itself.
We discuss here to what extent single-valued is this representation. 
In fact
$(1)$ contains the evident ``gauge freedom'':
$$\varphi = \varphi+ \omega, \quad (\triangle \omega=0), \qquad \mbox{\bf A}=\mbox{\bf A}+
\mbox{\boldmath $\nabla$} 
w.$$
If we remove the demands of topological triviality $\Omega$ or/and
the boundedness of functions 
$\omega$ and $w$, 
then as an example important for physical application, we may represent the 
gauge freedoms in the form of special additional functions to $\mbox{\bf V}$:
$$0\neq\mbox{\bf V}_{\mbox{\bf N}}=\mbox{curl}\mbox{\bf L} {{r^l}\choose {1/r^{l+1}}} Y_{lm} 
\equiv
{{-(l+1)\mbox{\boldmath $\nabla$} 
r^l Y_{lm}}\choose {l\mbox{\boldmath $\nabla$} r^{-l-1}Y_{lm}}}, $$
which have a nonzero finite value all over the space except $r \to \infty$ and
$r \to 0$, respectively, as far as   
$$ \mbox{div}\mbox{\bf V}_{\mbox{\bf N}}= \mbox{curl} \mbox{\bf V}_{\mbox{\bf N}}\equiv 0,  \quad
\mbox{in} \quad R^3 / S_{r \to \infty}^{2} \quad \mbox{and}\quad 
R^3/ \{0\}.$$
Therefore these functions~\footnote{Note that similar functions may be 
generated using the common relation
$$\mbox{curl}(\mbox{\bf r}\times \mbox{\boldmath 
$\nabla$}) r^{\kappa}Y_{lm}=
-(\kappa +1)\mbox{\boldmath 
$\nabla$} r^{\kappa}Y_{lm}+(\kappa -l)(\kappa+l+1)\mbox{\bf r} r^
{\kappa -2}Y_{lm},$$
and taking into account that
$$\kappa =l \ \ \mbox{and} \ \ \kappa=-l-1$$
(see \cite{Af}, Appendix A).} 
cannot be represented by the usual Helmholtz 
decomposition, and manifest the gauge freedom of its enlarged formulation. 

{\bf\small Remark I.} {\small Note that functions $\mbox{\bf V}_{N}$ are 
longitudinal and transverse simultaneously, since they represent the 
vector solutions of the Laplace equation ($\triangle \equiv 
\mbox{grad} \, \mbox{div} - \mbox{curl} \, \mbox{curl}$). 
Moreover, in our context it is important to emphsize that 
they are topologically equivalent to the poloidal (meridional) harmonics on 
the toruslike surface covering the whole space of $R^3$ 
except one deleted axis. }

Thus, under this gauge freedom the Helmholtz decomposition takes the
following alternative forms 
 $$ \mbox{\bf V}\equiv \mbox{\bf V}^{\|}+\mbox{\bf V}^{\bot}= $$
\begin{eqnarray}
= \mbox{\boldmath $\nabla$} \left( \triangle^{-1}\mbox{div} \mbox{\bf V} - (l+1) \sum_{lm} \bigl[
C_{lm} r^l + C_{lm}^{\prime} r^{-l-1}\bigr] Y_{lm} \right)
+\mbox{curl} \triangle^{-1}\mbox{curl}\mbox{\bf V}=  
\label{N3} \\
=  \mbox{\boldmath $\nabla$} \triangle^{-1}\mbox{div} \mbox{\bf V}+ \mbox{curl} \left( \triangle^{-1}\mbox{curl}\mbox{\bf V}+  \mbox{\bf L} \sum_{lm} \bigl[
C_{lm} r^l + C_{lm}^{\prime}  r^{-l-1}\bigr] Y_{lm} \right).
\nonumber
\end{eqnarray}

\section{The Neumann-Debye decomposition}

$\quad$The well-known mathematical physicist W.M.Elsasser 
in \cite{El} has already observed that every vector field of the
form $\mbox{\boldmath $\nabla$}\times \mbox{\boldmath $\nabla$}\times \mbox{\bf r}\chi +\mbox{\boldmath $\nabla$}\times
\mbox{\bf r} \psi$, where $\psi$ and $\chi$ are any scalars, is solenoidal. In 
paper \cite{Bs} it has been shown that if $\mbox{div} \mbox{\bf V} =0$ in $R^3$, then for every 
choice of the origin there exist unique scalars $\psi$ and $\chi$ such that
$\mbox{\bf V}=\mbox{\bf L} \psi +\mbox{\bf N} \chi$ while $\psi$ and $\chi$ average to zero
on every spherical surface concentric with the origin. The complete theorem
of the possibility of decomposition of $\mbox{\bf V}(\mbox{\bf r})$ in terms of scalar
functions reads as follows:

{\large Representation Theorem.}
Given a region $\Omega \subseteq R^3\backslash \{0\},$ with a regular boundary
and $R^3-$vector field, $\mbox{\bf V}:\ \mbox{\bf r}\in \Omega \rightarrow \mbox{\bf V}(\mbox{\bf r})
\in R^3.$ Then, there exist three scalar functions $\varphi (\mbox{\bf r}),
\psi (\mbox{\bf r}) \ \mbox{and} \ \chi (\mbox{\bf r}) \ \mbox{on} \ \Omega$ which
define this $\mbox{\bf V}$ \cite{Mul}. 
\special{em:linewidth 0.4pt}
\unitlength 1mm
\linethickness{0.4pt}
\begin{picture}(00,00)
\put(00,00){\rule{1.67\unitlength}{1.67\unitlength}} 
\end{picture}

The most used decomposition, which we call the Neumann-Debye one, has the
form
\begin{equation}
\triangleleft \qquad  \mbox{\bf V}(\mbox{\bf r}):=
\mbox{\boldmath $\nabla$}
\varphi(\cdot)+\mbox{curl}\mbox{\bf r}\psi(\cdot)+
\mbox{curl} \ \mbox{curl}\mbox{\bf r}\chi(\cdot) \equiv 
\mbox{\boldmath $\nabla$} \varphi +\mbox{\bf L}\psi +
\mbox{\bf N}\chi .
\end{equation}
Here $\psi$ and $\chi$ are the so-called Debye potentials and $\varphi$
is the usual (electric) scalar one. 
We found
fundamental solutions of the inversion problem of (1) in the form~\cite{DM}
\begin{eqnarray}
\mbox{div} \mbox{\bf V}=\triangle \varphi \qquad \rightarrow \qquad
\varphi&=&\triangle^{-1}\mbox{div} \mbox{\bf V},  \nonumber\\
\mbox{\bf L}\mbox{\bf V} = L^2\psi  \qquad \rightarrow \qquad
\psi&=& - L^{-2}\mbox{\bf L}\mbox{\bf V}\equiv \mbox{\bf L}^{-2}\mbox{\bf r}\  \mbox{curl}\mbox{\bf V}, \label{4} \\
\mbox{\bf r}\mbox{\bf V}=(\mbox{\bf r}\mbox{\boldmath $\nabla$} )
\varphi +L^2\chi \qquad \rightarrow \qquad
\chi&=& L^{-2}(\mbox{\bf r}\mbox{\boldmath $\nabla$})
\triangle^{-1} \mbox{div}\mbox{\bf V} -
L^{-2}(\mbox{\bf r}\mbox{\bf V}),  \nonumber 
\end{eqnarray}
where \cite{Cou}
\begin{displaymath}
\mbox{\bf L}^{-2}:=\int_{\sigma}{{\frac{d\omega'}{4\pi}}ln(1-\hat r\cdot
\hat r')}. \quad  \triangleright
\end{displaymath}

{\bf\small Remark II.} {\small Eigenfunctions of the square of the
operator angular momentum $iL:\ (iL)^2=-L^2$ are usual spherical
functions satisfying the equation
\begin{eqnarray}
L^2\ Y_{lm}(\hat r)=-l(l+1)Y_{lm}(\hat r) \nonumber
\end{eqnarray}
The corresponding Green function for this equation can be found with the 
help of the known Mercer theorem (\cite{Cou}, v.1) which in the given case 
yields}
\begin{eqnarray}
\frac{1}{4\pi}\sum_{l,m}\frac{Y_{lm}^*(\hat r)Y_{lm}(r')}{-l(l+1)}=
-\sum_{l}\frac{2l+1}{l(l+1)}P_l(\hat r\cdot \hat r')=1-ln2+ln(1-\hat r\cdot
\hat r'). \quad  
\special{em:linewidth 0.4pt}
\unitlength 1mm
\linethickness{0.4pt}
\begin{picture}(00,00)
\put(00,00){\rule{1.67\unitlength}{1.67\unitlength}} 
\end{picture}
\nonumber
\end{eqnarray}

{\bf\small Remark III.} {\small It is well-known that the gauge freedom of 
(\ref{4}) is the following:
$$\varphi \rightarrow \varphi +C, \quad  \psi \rightarrow \psi +\mu (r),
\quad \chi \rightarrow \chi +\nu (r).$$
Requiring $\varphi$ to vanish on the boundary, and $\psi, \ \chi$ not to
contain spherically symmetric components
$$\int_{S^2}{dw \ \psi}\ = \ \int_{S^2}{dw\ \chi} \ =\ 0; $$
we put these functions in one-to-one correspondence to $\mbox{\bf V}$.}
\special{em:linewidth 0.4pt}
\unitlength 1mm
\linethickness{0.4pt}
\begin{picture}(00,00)
\put(00,00){\rule{1.67\unitlength}{1.67\unitlength}} 
\end{picture}

{\large Uniqueness Theorem.} If a vector field $\mbox{\bf V}$ (with the
properties determined in the theorem (I) ) is defined on every $S_r^2$ in
some range $r_0<r<r_1$ and in that range $V_r=0$ while $V_{\theta}(r,\theta ,
 \varphi)$
and $V_{\varphi}(r,\theta,\varphi)$ are bounded for each fixed $r$ and are
continuously differentiable except possibly at $\theta=0$ and $\theta=\pi$
and  further $\mbox{div} \mbox{\bf V}=0$ and $\mbox{curl} \mbox{\bf V}=0$, then $\mbox{\bf V}\equiv 0$.

Our inversion formulas (3) demonstrate that immediately 
(cp. with \cite{Bs}, p.383, where the condition $\mbox{\bf L}\mbox{\bf V}=0$ has been 
used instead of our $\mbox{curl} \mbox{\bf V}=0$). 
\special{em:linewidth 0.4pt}
\unitlength 1mm
\linethickness{0.4pt}
\begin{picture}(00,00)
\put(00,00){\rule{1.67\unitlength}{1.67\unitlength}} 
\end{picture}

Now we have to reconstruct the representation of $\chi$ such that 
it depends on $\mbox{curl} \mbox{\bf V}$ and $\mbox{div} \mbox{\bf V}$ only. Really, the
latter quantities have the physical meaning but not the radial component of
$\mbox{\bf V}$. Moreover, we may expect $\chi$ not to depend on 
$\mbox{div} \mbox{\bf V}$ generally because this potential defines the transverse
part of the vector field $\mbox{\bf V}$. Nevertheless, because of gauge freedom, 
the situation is not so simple as it seems to be. 

Indeed, we may substitute the Helmholtz
decomposition (the last expression in (\ref{N3})) into the term with 
$\mbox{\bf r} \mbox{\bf V}$ and see that $\chi$ takes the form 
\begin{eqnarray}
\chi &=& L^{-2} \mbox{\bf L} \triangle^{-1} \mbox{curl} \mbox{\bf V} 
\sum_{lm} L^{-2} \bigl[
C_{lm} \nabla r^l + C_{lm}^{\prime} \nabla r^{-l-1}\bigr] Y_{lm}.
\end{eqnarray}
Further our vector field corresponding to the gauge freedom may be 
transformed as
\begin{eqnarray}
 \mbox{curl} \mbox{\bf L} L^{-2} (r \nabla) r^{l}Y_{lm}=
(l+1) \mbox{curl} r^{l} \mbox{\bf L} L^{-2} Y_{lm}=  \nonumber \\
 = (l+1) \mbox{curl} r^{l} \mbox{\bf L} L^{-2} \frac{L^2}{l(l+1)}Y_{lm}=
\frac{1}{l}\mbox{curl} \mbox{\bf L} r^{l} Y_{lm}=
-\frac{l+1}{l} \mbox{\boldmath $\nabla$} r^{l}Y_{lm}.
\label{N7}
\end{eqnarray}

So, we found fundamental solutions of the inversion problem of (\ref{4}) 
in the form~\cite{DM}
\begin{eqnarray}
\mbox{div} \mbox{\bf V}=\triangle \varphi \qquad \rightarrow \qquad
\varphi&=&\triangle^{-1}\mbox{div} \mbox{\bf V},  \nonumber\\
\mbox{\bf L}\mbox{\bf V} = L^2\psi  \qquad \rightarrow \qquad
\psi&=& - L^{-2}\mbox{\bf L}\mbox{\bf V}\equiv \mbox{\bf L}^{-2}\mbox{\bf r}\  \mbox{curl}\mbox{\bf V}, \label{N6} \\
\mbox{\bf r}\mbox{\bf V}=(\mbox{\bf r}\mbox{\boldmath $\nabla$}
 )\varphi +L^2\chi \qquad \rightarrow \qquad
\chi&=& L^{-2}(\mbox{\bf r}\mbox{\boldmath $\nabla$})\triangle^{-1} \mbox{div}\mbox{\bf V} -
L^{-2}(\mbox{\bf r}\mbox{\bf V}),  \nonumber \\
&=& L^{-2} \mbox{\bf L} \triangle^{-1} \mbox{curl} \mbox{\bf V} \nonumber \\
&+&  \sum_{lm} L^{-2} \bigl[
C_{lm} \nabla r^l + C_{lm}^{\prime} \nabla r^{-l-1}\bigr] Y_{lm} \nonumber
\end{eqnarray}
with taking into account the gauge freedom in (\ref{N3}). $\quad $
\special{em:linewidth 0.4pt}
\unitlength 1mm
\linethickness{0.4pt}
\begin{picture}(00,00)
\put(00,00){\rule{1.67\unitlength}{1.67\unitlength}} 
\end{picture}

\section{The multipole expansion of the \\
electromagnetic current}

$\quad$Now we compare our abstract exercises and the 
procedure of 
multipole expansion of electromagnetic current  
on the basis of the vector Helmholtz equation solutions constructed through
the Neumann-Debye decomposition (\ref{N6}). 
The multipole representation of current $\mbox{\bf J}(\mbox{\bf r})$ by (\ref{4}) may be 
obtained by the standard
expansion of three scalar densities into series:
\begin{eqnarray}
\phi \sim \sum_{lmk} j_{l}(kr) Y_{lm} \dot{Q}_{lm} (k^2, t); 
\nonumber \\
\psi \sim \sum_{lmk} j_{l}(kr) Y_{lm} M_{lm} (k^2, t); 
\label{N9} \\
\chi \sim \sum_{lmk} j_{l}(kr) Y_{lm} E_{lm} (k^2, t).
\nonumber
\end{eqnarray}

Therefore the multipole representations of the transverse
part of current ($\mbox{div} \, \mbox{\bf J} =0$) are determined by the magnetic
form factors $M_{lm}(k^2, \, t)$, and the transverse electric contributions
$E_{lm}(k^2, \, t)$, and the scalar part of current ($\mbox{curl} \, \mbox{\bf J} =0$)
are expressed due to the $4-$current conservation law ($\mbox{div} \, \mbox{\bf J} =
- \dot{\rho}$) through the Coulomb (charge) multipole moments $Q_{lm}(0 \, t)$
and their mean $2n-$power radii
$$ Q_{lm}(k^2, \, t) = Q_{lm} (0, \, t) + \sum_{n=1}^{\infty} 
\frac{k^{2n}}{n!} Q_{lm}^{(2n)} (0, \, t).$$
As a result, $\mbox{\bf J}$ may be represented as \cite{DbCh}
\begin{eqnarray}
\mbox{\bf J}(\mbox{\bf r}, \, t)& &= (2\pi)^{-3} \sum_{l,m} \int_{0}^{\infty} dk  
(-ik)^{l} \frac{\sqrt{4 \pi (2l+1)}}{l(2l+1)!!} 
\{ -  i k \, \mbox{\bf L} f_{l}(kr)Y_{lm}(\hat{\mbox{\bf r}})
M_{lm}(k^2, \, t)  \nonumber \\
& & -ik \mbox{curl} \mbox{\bf L}   
f_{l}(kr)Y_{lm}(\hat{\mbox{\bf r}}) 
E_{lm}(k^2, \, t) +  
l \, \mbox{\boldmath $\nabla$} f_{l}(kr)Y_{lm}(\hat{\mbox{\bf r}}) \dot{Q}_{lm}(k^2, \, t)  \}. 
\label{N10}
\end{eqnarray}

We repeat here the procedure used firstly in \cite{DbCh} for the
ascertainment of the exact structure of transverse electric 
contributions $E_{lm}(k^2, \, t)$. To this end, it one should rewrite
explicitly our basis functions in terms of the vector harmonics
\begin{eqnarray}
\mbox{curl} \mbox{\bf L} f_{l}(kr)Y_{lm}(\hat{\mbox{\bf r}}) = (2l+1)^{-1/2} \{ f_{l-1}
(kr) \sqrt{l+1} \mbox{\bf Y}_{ll-1m}(\hat{\mbox{\bf r}}) + 
\nonumber \\
+ f_{l+1}(kr) \sqrt{l} \mbox{\bf Y}_{ll+1m}(\hat{\mbox{\bf r}}) \};
\nonumber 
\end{eqnarray}
\begin{eqnarray}
\mbox{\boldmath $\nabla$} f_{l}(kr)Y_{lm}(\hat{\mbox{\bf r}}) = (2l+1)^{-1/2} \{ f_{l-1}
(kr) \sqrt{l} \mbox{\bf Y}_{ll-1m}(\hat{\mbox{\bf r}}) -
\nonumber \\
- f_{l+1}(kr) \sqrt{l+1} \mbox{\bf Y}_{ll+1m}(\hat{\mbox{\bf r}}) \},
\nonumber 
\end{eqnarray}
where $\mbox{\bf Y}_{ll-1m}(\hat{\mbox{\bf r}})$ is a harmonic polynomial function defined as
$$ r^{l-1}\mbox{\bf Y}_{ll-1m}(\hat{\mbox{\bf r}}) = \frac{1}{\sqrt{l(2l+1)}} 
\mbox{\boldmath $\nabla$
} r^{l} Y_{lm}.$$
As is obvious, in the wave-length approximation, the leading contributions in the 
latter expressions are delivered by the vector harmonic functions
$f_{l-1}Y_{ll-1m}$ from which it follows that
\begin{eqnarray}
\mbox{curl} \mbox{\bf L} f_{l}(kr)Y_{lm}(\hat{\mbox{\bf r}}) \approx_{k \rightarrow 0} 
\sqrt{(l+1)/l} \mbox{\boldmath $\nabla$} f_{l}(kr)Y_{lm}(\hat{\mbox{\bf r}})=
\nonumber  \\
=\frac{4\pi (ikr)^{l-1}}{(2l+1)!!} \sqrt{\frac{l+1}{l}}
\mbox{\boldmath $\nabla$} r^{l} Y_{lm}(\hat{\mbox{\bf r}})+ O[(kr)^{l+1}]
\label{N11}
\end{eqnarray}
It is the relation that permits us to identify the leading term in 
$E_{lm}(k^2, \, t)$ with time-derivatives of $Q_{lm}(0, \, t)$.
However all functually independent contributions in $E_{lm}(k^2, \, t)$
give the so-called toroid moments and their $2n-$power radii
$$ T_{lm}^{(2n)}(0, \, t) = - \frac{\sqrt{\pi l}}{2l+1} \int r^{l+2n+1}
\left[ \mbox{\bf Y}_{ll-1m}^{*} (\mbox{\bf r}) + \frac{2\sqrt{l/(l+1)}}{2l+3} \mbox{\bf Y}_{ll+1m}^{*}(\mbox{\bf r})
\right] \mbox{\bf J} (r, t) d^3 r.$$
As the toroid moments have a distinct geometrical meaning
(diverse details and representations are given in \cite{DbCh}, \cite{Vlad}, 
\cite{DubTug}, \cite{Shab} and see also \cite{Saha}), the rejection of
$T_{lm}(t)$ is generally invalid like it was done in the Siegert theorem
$E_{lm} (k^2, \, t) \rightarrow_{k \rightarrow 0} \dot{Q}_{lm} (0, \, t)$.
Neglect of $T_{lm}(t)$ in comparison with $Q_{lm}(t)$ is analogous to the
neglect of a higher multipole moment (contributions of highest symmetries
of a given system) in comparison with the lower ones, which is of course 
permissible only when the lower moments (low symmetries) of this system 
do exist. So, the strict theorem determining the electric part structure 
has the form (\ref{N1})
$E_{lm}(\mbox{\bf k}, \, t) = \dot{Q}_{lm}(0, \, t) + \mbox{\bf k}^2 T_{lm} (\mbox{\bf k}, \, t),$
and its validity and uniqueness rely on the
gauge freedom which has been obtained for the enlarged Helmholtz theorem and
transferred to  
Debye potentials (compare (\ref{N6}) and (\ref{N11})).

Moreover, by using the exact relation (\ref{N1}),  
it is not so hard to find an 
expression for the complete parametrization of the current in 
terms of generalized functions~\footnote{The contribution of $l=0, \, n=0$ to the last term is 
forbidden by total charge conservation, whereas other terms contain 
no contribution of $l=0$, formally, owing to 
$(\mbox{\bf r} \times \mbox{\boldmath $\nabla$}) \delta(\mbox{\bf r}) \equiv 0$.} \cite{DubTug}
\begin{eqnarray}
\label{N12}
\mbox{\bf J} (\mbox{\bf r}, \, t) =  \sum_{l=1}^{\infty} \sum_{m=-l}^{m=l} \sum_{n=0}^{\infty}
\frac{(2l+1)!!}{2^n n! l(2l+2n+1)!!} \sqrt{\frac{4 \pi}{2l+1}}
\{  M_{lm}^{(2n)} (t) (\mbox{\bf r} \times \mbox{\boldmath $\nabla$}) \Delta^{n} \delta_{lm}(\mbox{\bf r}) \nonumber\\
\\
+  \bigl[ \dot{Q}_{lm} (t)  \delta_{n,0} \Delta^{-1} - T_{lm}^{(2n)} (t)
\bigr] \mbox{curl} (\mbox{\bf r} \times \mbox{\boldmath $\nabla$}) \Delta^{n} \delta_{lm} (\mbox{\bf r}) - 
l \, \dot{Q}_{lm}^{(2n)} (t) \mbox{\boldmath $\nabla$} \Delta^{n-1} \delta_{lm}(\mbox{\bf r})\}.\nonumber
\end{eqnarray}

{\bf\small Remark IV.} {\small  However, the expansion via of spherical
harmonics emerges rapidly convergent. We remind (see e.g. \cite{Bl}, p.806) 
that if the field decreases for large distances very slowly, slower than
$r^{-2}$, the divergence and curl of the vector field considered are assigned 
arbitrary independent values. Conversely, if we know that $\mbox{\bf J}$ vanishes 
identically outside some source radius $R$, $\mbox{\boldmath $\nabla$} 
\mbox{\bf J}$ and $\mbox{\boldmath $\nabla$} \times \mbox{\bf J}$ 
are no longer independent of each other. As far as the expansion via 
spherical harmonics is rapidly convergent, it is realized in the latter
representation (\ref{N12}) immediately.}

Thus, we strictly demonstrated that the gauge freedom in division
of the electromagnetic current into the transverse and longitudinal parts
leads to the fact that the multipole contributions to the 
transverse part of current $E_{lm} (k^2, t)$ are represented in the form 
(\ref{N1}) and its leading terms 
may be identified with $\dot{Q}_{lm} (0 \, t)$ from the 
longitudinal part of current for all $l$. Note that, since the coefficient 
$C_l$ does not depend on the wave number $k$, we can use hereafter the 
Lorentz gauge condition in the calculation of the vector potential. 
\section{ Conclusion}

$\quad$The representation of $\chi$ in the Neumann-Debye 
scalarization already assumes that the prohibition of the electric type of 
radiation imposes some conditions both on ${\rm curl} \mbox{\bf J}$ and ${\rm div} \mbox{\bf J}$.
But we could not reveal the ones due to their non-division in the 
scalarization mentioned. 
Exploitation of the enlarged Helmholtz theorem for this operation has
inserted the extended gauge freedom (let us recall that the Neumann-Debye 
representation the gauge freedom reduce to functions of the scalar
argument $|\mbox{\bf r}|$ only). It is just this freedom, consideration of which made it
possible to identify the coefficients of leading order of the expansion
of transverse and longitudinal electric parts of the current!

The form of the expression (\ref{N1}) shows the possibility of compensation
of the electric type radiation if the toroid and charge moments are switched on
as ``anti-phase'' ones~\cite{DbCh} (see also \cite{Afan}).

\end{document}